# A Simplified Approach to Simulating Raman Spectra from *Ab Initio* Molecular Dynamics

*Edoardo Aprà,[1] Ashish Bhattarai,[2] Eric Baxter,[2] Grant E. Johnson,[2] Niranjan Govind,[1] and Patrick Z. El-Khoury,[2,*]*

[1]Environmental and Molecular Sciences Laboratory

[2]Physical Sciences Division

Pacific Northwest National Laboratory, P.O. Box 999, Richland, WA 99354, USA.

[*]*patrick.elkhoury@pnnl.gov*

**ABSTRACT**

We describe a simplified approach to simulating Raman spectra using *ab initio* molecular dynamics (AIMD) calculations. Our protocol relies on on-the-fly calculations of approximate molecular polarizabilities using a sum over orbitals (as opposed to states) method. This approach greatly speeds up the simulation time (~8X/step on average for functionalized aromatic thiols) by bypassing the nominally more accurate but computationally expensive approach to calculating molecular polarizability, i.e., solving the coupled perturbed Hartree-Fock/Kohn–Sham equations. The speedup is dramatic when some 100K on-the-fly polarizability calculations are required to converge simulated AIMD-Raman spectra to their experimental analogues, particularly for large molecular systems. We demonstrate the advantages and limitations of our method through a few case studies targeting molecular systems of interest to on-going experimental efforts.

**Introduction**

There are numerous advantages to simulating Raman spectra using *ab initio* molecular dynamics (AIMD) simulations.[1,2,3] This is the case for both conventional calculations aimed at recovering ensemble-averaged molecular spectra[1-3] as well as more recent applications in the areas of surface-enhanced[4,5] and single molecule[6] Raman scattering. That said, AIMD-based simulation of Raman spectra is relatively computationally expensive.[7] For instance, we recently showed that converging the simulated Raman line shapes of a small molecular system (dimethyl sulfoxide) to their experimental analogues requires a total simulation time of over 200 ps and 400K molecular polarizability calculations.[3] The number of calculations becomes increasingly prohibitive for medium and large molecular systems, where proper conformational sampling necessitates even longer simulation times. Although relatively long propagation times may be circumvented by performing and averaging over the results of several short trajectories,[3,6,8] molecular polarizability calculations continue to be the bottleneck limiting AIMD-based Raman spectral simulations. This challenge motivates our current work.

Molecular polarizabilities may be conveniently computed by solving the coupled perturbed Hartree–Fock or Kohn–Sham equations (CPHF/CPKS).[9,10,11,12,13,14,15,16,17,18,19] Sum over states (SOS) formalisms[20,21] within the framework of time-dependent density functional theory (TD-DFT) are used less often, principally because converging polarizabilities requires computing the transition energies and oscillator strengths for a large number of electronic states.[22] Within the so-called Dalgarno uncoupling formalism scheme, SOS equations may be regarded as uncoupled self-consistent field equations.[23] The resulting formulae involve sums over occupied and virtual molecular orbitals rather than sums over excited electronic states, which is why they have been appropriately termed sum over orbitals (SOO) equations.[24] The SOO method relies on several approximations, the most detrimental of which is neglecting the coupling portion of the perturbation and replacing state energies and oscillator strengths by orbital energy differences and inter-orbital transition strengths.[23,24] In other words, both the Coulomb and exchange integrals representing the self-consistent field correction to the orbital energies as well as additional energy corrections due to the perturbation are entirely neglected. Nevertheless, a previous study that employed the SOO method to compute polarizabilities and hyperpolarizabilities of benzene, nitrobenzene, phenols, and nitro-derivatives of phenol found this technique to be satisfactory.[24] Indeed, the simplified approach yielded trends that agree with their experimental analogues at dramatically reduced computational cost.

Notwithstanding the absolute accuracy of the SOO polarizabilities, the question raised herein may be distilled into: are time variations about the calculated average polarizabilities in the AIMD scheme physically meaningful? Our rather strict gauge of the latter is the quality of AIMD Raman spectra derived from Fourier transforms of polarizability autocorrelation functions.[1-3] Recall that approximations at the level of molecular polarizability calculations are expected to alter the relative intensities of the Raman-active vibrational states when compared to more accurate approaches to computing on-the-fly polarizabilities; vibrational resonances have to do with the level of theory and conformational sampling in the AIMD scheme, and hence, should be minimally affected. With this in mind, our original goal was to identify Raman-active vibrational modes in the resulting AIMD vibrational spectra, akin to a vibrational density of states plot filtered by Raman selection rules. To our surprise, we found the derived SOO and CPKS-based AIMD Raman spectra to be very similar in quality when compared to experimentally measured spectra. We demonstrate the latter using several molecular systems of interest to ongoing experimental work, namely, thiophenol (TP), 4-mercaptobenzonitrile (MBN), 4-nitrothiophenol (NTP), 4-aminothiophenol (ATP), 4,4'-dimercaptoazobenzene (DMAB), and benzene (for reference). The following section describes the theoretical and computational approaches adopted herein.

**Methods**

**Formalism.** The familiar SOS formula that may be used to compute the $AB^{th}$ component of the polarizability tensor is given by[20-24,25]

$$\alpha_{AB}(-\omega;\omega) = \sum_{i\neq 0}\left[\frac{\mu^A_{0i}\mu^B_{i0}}{\Delta_i - \omega} + \frac{\mu^B_{0i}\mu^A_{i0}}{\Delta_i + \omega}\right] = \hat{P}[A(-\omega), B(\omega)]\sum_{i\neq 0}\frac{\mu^A_{0i}\mu^B_{i0}}{\Delta_i - \omega}$$

in which

$$\mu^A_{ij} = \langle i|\hat{\mu}|j\rangle$$

$A, B$ denote the x, y, and z directions, and A = B corresponds to a diagonal polarizability element; $\omega$ is the energy of the external field and setting $\omega = 0$ recovers the static polarizability; $\Delta_i$ is the energy difference between the ground ($0^{th}$) and $i^{th}$ excited state; $\hat{P}$ is a permutation operator, which in the case of $\alpha = 2!$;[24,25] $\mu^A_{ij}$ is the $A^{th}$ component of the transition dipole between states $i$ and j, and when $i = j$, the term collapses to the electric dipole moment of the state; $\hat{\mu}$ is a dipole moment operator. Within the SOO approximation, the quantities of interest are the products of electric dipole transition moments between occupied and virtual molecular orbitals (numerator) and the energy differences between the molecular orbital energies and the energy of the perturbing optical field (denominator). Namely, within the SOO framework and in the static limit, the $AB^{th}$ component of the molecular polarizability is reduced to[24,25]

$$\alpha_{AB} = 2\hat{P}\sum_{o,v} n_i \frac{\langle\varphi_o|\hat{\mu}_B|\varphi_v\rangle\langle\varphi_v|\hat{\mu}_A|\varphi_o\rangle}{\varepsilon_v - \varepsilon_o}$$

Here, $n_i$ is the occupation number ($0 \leq n_i \leq 1$ for open shell and $0 \leq n_i \leq 2$ for closed shell); the subscripts $o$ and $v$ denote occupied and virtual orbitals $\varphi$ with energies $\varepsilon$. We use this formula to compute molecular polarizabilities on-the fly (see below) and account for the transitions between all molecular orbitals throughout. Accurate molecular polarizabilities are otherwise computed for reference using the more accurate CPKS method, as implemented in NWChem.[26,27,28]

Raman scattering activity ($S_m$) is given by[29,30]

$$S_m = g_m[45{\alpha'_m}^2 + 7{\beta'_m}^2]$$

in which

$$\alpha'_m = \frac{1}{3}(\tilde{\alpha}'_{xx,m} + \tilde{\alpha}'_{yy,m} + \tilde{\alpha}'_{zz,m})$$

and

$${\beta'_m}^2 = \frac{1}{2}[(\tilde{\alpha}'_{xx,m} - \tilde{\alpha}'_{yy,m})^2 + (\tilde{\alpha}'_{yy,m} - \tilde{\alpha}'_{zz,m})^2 + (\tilde{\alpha}'_{zz,m} - \tilde{\alpha}'_{xx,m})^2 + 6({\alpha'_{xy,m}}^2 + {\alpha'_{xz,m}}^2 + {\alpha'_{yz,m}}^2)]$$

In the above equations, $g_m$ is the degeneracy of the vibrational states and primes denote derivatives with respect to the $m^{th}$ vibrational state, $\alpha'_m/{\beta'_m}^2$ are the isotropic/anisotropic polarizabilies, and $\tilde{\alpha}'_{ij,m}$ ($i,j = x,y,z$) are elements of the 3 x 3 polarizability derivative tensor. Regardless of how polarizabilities are evaluated, Raman spectra are obtained from Fourier transforms of the (averaged) polarizability autocorrelation functions, as described in prior works.[1-3] Note that no frequency or intensity correction factors are otherwise used in this study.

**Computational.** All calculations were performed using a local development version of NWChem,[31] with the PBE exchange-correlation functional[32] in conjunction with the def2-TZVP basis set.[33] We will show that this level of theory recovers the experimental observables. Our choice of starting structures consist of the fully optimized molecular structures for all systems. All of the atoms were otherwise relaxed in the subsequent AIMD trajectory calculations.[3] A time step of ~0.5 fs was used throughout the trajectory calculations, such that the total energies of the systems were conserved to within 1-2 kcal/mol. To generate the initial conditions for subsequent constant energy simulations, we first ran 10 ps-long trajectories (one for each system) in the canonical ensemble using the stochastic velocity rescaling thermostat of Bussi *et al.*,[34] with a relaxation parameter of 100 atomic units. After equilibration, randomly selected structures from these initial trajectories were used for the production runs. For each molecular system, a total of at least 10 trajectory calculations (over 5 ps total simulation time/trajectory) were then performed in the microcanonical ensemble and concatenated to yield the ensemble averaged optical spectra.

**Experimental.** High-resolution bulk Raman spectra from pure TP (Sigma-Aldrich), MBN crystals (Aurum Pharmatech), NTP (Aurum Pharmatech), ATP (Tokyo Chemical Industry), and benzene (Sigma-Aldrich, measured for reference) were recorded using an inverted optical microscope (Nikon Ti-E) coupled to a Raman spectrometer (Horiba LabRAM HR). Micro-Raman measurements were performed using a 633 nm laser, which was attenuated using a variable neutral density filter wheel (to 25 $\mu W/\mu m^2$), reflected off a dichroic beam splitter, and focused onto the sample using a 60x/0.7 numerical aperture (NA) air objective. The backscattered light was collected through the same objective, transmitted though the beam splitter cube, and dispersed through an 1800 g/mm grating onto a CCD camera.

DMAB was produced through the dimerization of NTP or ATP molecules. In the case of NTP, DMAB was formed at a plasmonic tip-sample nanojunction in aqueous solution, as described in a recent work from our group.[35] DMAB was otherwise formed throughout the course of electrospray deposition (ESD[36,37]) of an ethanolic solution of ATP (3.7 mM) onto an indium tin oxide (ITO) substrate. Solutions were infused through a pulled fused silica capillary (∼75 μm inner diameter) tip at a flow rate of 50 μL hour$^{-1}$. A 1 mL glass syringe (Hamilton) mounted on a microprocessor-controlled syringe pump (KD Scientific) was used to infuse the solution through the electrospray capillary. After 1 hour of deposition (50 μL total volume) using the positive ionization mode (+4 kV), an area (around 1 cm$^2$) of adsorbed species was faintly visible by the naked eye at the surface of the electrically grounded ITO substrate. Raman spectra were then recorded using a previously described setup.[38,39] For the purpose of this report, a 633 nm laser (100 - 200 μW) was incident onto the sample through a 100X air objective (Mitutoyo, NA = 0.7) using our top excitation channel (angle of incidence normal to the sample surface). The scattered radiation was collected through the same objective and filtered through dichroic and long pass filters. The resulting radiation was detected by a CCD camera (Andor, Newton EMCCD) coupled to a spectrometer (Andor, Shamrock 500, 300 l/mm grating).

**Results and Discussion**

It has been previously shown that uncoupled Hartree-Fock and Kohn-Sham approximations introduce significant errors in the derived polarizability values.[24,25,40,41] Comparing CPKS and SOO polarizabilities for the aromatic thiols considered herein, therefore, constitutes a good starting point for our discussion, see Figure 1. Following full geometry optimization of benzene, TP, ATP, NTP, MBN, and DMAB, CPKS and SOO polarizabilities were computed at the minima, see Figure 1A. We find that the derived isotropic and anisotropic SOO polarizabilities are overestimated (~2X on average) with respect to their CPKS analogues for all of the molecular systems considered in this analysis. Besides a few noticeable outliers (anisotropic polarizabilities of TP and NTP), a linear trend between the SOO and CPKS values is observed, which is indicative of the systematic nature of the error introduced by the SOO approximation. Interestingly, we find improved correlation between time-averaged SOO and CPKS polarizabilities, obtained by averaging the values computed using the two methods at every time step throughout the course of trajectory calculations (the same runs used to evaluate Raman spectra using the two methods, see below). We present our case using the two systems that exhibit the largest deviations in our above-described analysis, namely TP and NTP, as well as MBN for reference, see Figure 1B. As expected, the absolute values of the time-averaged SOO polarizabilities are still larger than their CPKS analogues. Nonetheless, the linear correlation is significantly improved ($R^2 = 0.98$). As discussed below, the same conclusion can be drawn on the basis of the computed AIMD Raman spectra.

Another approach to gauging the systematic error introduced by the SOO approximation is to compare AIMD Raman spectra obtained using SOO and CPKS polarizabilities computed on-the-fly throughout the course of trajectory calculations. To this end, we compare experimental, static/harmonic Raman spectra evaluated at minimum energy geometries, and AIMD Raman spectra derived using CPKS and SOO polarizabilities shown in Figure 2. The results for TP are compared in Figure 2A. The similarity between the two AIMD spectra is notable, both in terms of the frequencies and relative intensities of the observable vibrational states, particularly in the 750-1700 $cm^{-1}$ region of the spectrum. Note that the spectra were derived from two different trajectory runs, each of which spanned a total simulation time of ~50 ps. The largest deviation has to do with the line shape of the SH stretching mode (experimentally observed at 2571 $cm^{-1}$); the SOO-based spectrum predicts a much broader resonance with respect to the CPKS and experimental spectra (highlighted using a red rectangle in Figure 2A). The latter may be attributed to incomplete conformational sampling along the SOO trajectories. Other spectral features that are highlighted using yellow rectangles (resonance frequencies) and dashed lines (relative intensities, using the experimental values as reference) in the same figure illustrate that the predicted AIMD resonances and relative intensities (i) are similar when SOO and CPKS spectra are compared to one another, and (ii) reproduce the experimental values with much higher fidelity when compared to the static/harmonic spectrum that is computed at the energy minimum. The most notable deviations in terms of the resonance energies are for the CSH bending and SH stretching vibrations, experimentally at 918 and 2572 cm$^{-1}$, respectively. The two modes are underestimated by 59 and 132 cm$^{-1}$ using the static approach, whereas the AIMD values slightly underestimate the first (by 10 cm$^{-1}$) and otherwise overestimate the second (by 72 cm$^{-1}$) to a lesser extent.

The results for NTP are summarized in Figure 2B. In this case, the computed spectra all exhibit close overall resemblance to the experimental spectrum in the 750 - 1700 cm$^{-1}$ region. The red rectangle shown on the spectrum highlights a region where deviations between the AIMD and experimental spectra are evident. Namely, the (relative) intensity of the 859 cm$^{-1}$ mode, which can be assigned to $NO_2$ bending coupled to CH stretching, is well-captured using the harmonic approach, whereas it is overestimated using both CPKS and SOO-based AIMD Raman spectra. That both approaches yield similar deviations is indicative of the systematic nature of the error that is associated with the SOO approximation with respect to the CPKS approach. This is consistent with the results shown in Figure 1 and our above discussion. Towards the middle of the spectral range, small but noticeable vibrational signatures that are experimentally observed were only reproduced through the AIMD approach (see the ensuing figures and discussion). Conversely, the computed relative intensity of the SH stretching vibration is slightly overestimated using the harmonic approximation and relatively dim (barely visible) in the experimental and both AIMD spectra. The same is noticeable in the high frequency region of the Raman spectra of MBN, see Figure 2C. Furthermore, the AIMD spectra of MBN also better-capture the resonance frequency of the CN stretching vibration (2226/2236 cm$^{-1}$ experimental/AIMD resonances) when compared to the harmonic Raman spectrum (2176 cm$^{-1}$). Last but not least, the relative intensities of the vibrational states are again best captured using the AIMD approaches. This is consistent with prior observations from our group.[5] Interestingly, the spectrum computed using the SOO approximation exhibits closer resemblance to the experimental spectrum when compared to the AIMD spectrum computed using more exact (CPKS) polarizabilities. This may simply be the result of cancellation of error, but could be attributed to better convergence of the SOO trajectories that we reemphasize are distinct from the ones used to compute AIMD Raman spectrum within the CPKS framework (50 ps total simulation time for each case).

The plots shown in Figure 3 illustrate that the SOO-based AIMD Raman spectra are predictive and may be directly used to account for experimental observables. We illustrate the principle using the dimerization of ATP (Figure 3A) or NTP (Figure 3B) molecules to form DMAB as an example of current interest to on-going work.[35] Experimentally, this is achieved at a plasmonic tip-sample nanojunction following 633 nm laser irradiation for NTP (Figure 3C),[35] or throughout the course of electrospray deposition of ATP onto a ITO (Figure 3D). Assignments based on normal mode analysis as well as the (partial) vibrational density of states are shown in the supporting information. For the purpose of our discussion, we note that the overall agreement between the computed and experimental spectra, both in terms of the predicted resonance energies (a property of AIMD) as well as relative intensities of the vibrational states for all molecular systems considered in this analysis (governed by SOO polarizabilities), allows us to reliably distinguish between reactants and products.

Our next simulations take advantage of the reduced cost associated with SOO-based AIMD-Raman spectral simulations to tackle systems where CPKS calculations are difficult, or even prohibitively expensive. We also take advantage of Coulomb fitting,[42,43] which becomes very efficient when GGA functionals are used. The AIMD Raman spectra of a MBN-$Ag_{79}$ complex (panel A) and $C_{60}$ fullerene (panel B) is shown in Figure 4. In the former, we compare the spectra obtained from a relatively short AIMD trajectory (3.4 ps in length) to the experimental powder spectrum (black trace) and SOO-based AIMD spectrum of the isolated molecule (green trace). Given the short simulation time, which is necessitated by the large size of the system (1570 electrons), we did not expect the spectrum to be fully converged. Nonetheless, and with the exception of 1183 and 1203 $cm^{-1}$ vibrations that arise from C-C(N) stretching vibrations with various contributions from aromatic CH in-plane motions, the MBN-$Ag_{79}$ spectrum shows the familiar CN (2256 $cm^{-1}$), aromatic CC (1575 $cm^{-1}$), and C-S(H) (1057 $cm^{-1}$) stretching vibrations. Deviations of the computed molecule-metal complex vibrational signatures from there analogues for the isolated system deserve further scrutiny. The latter is beyond the scope of this work.

In the case of the fullerene, the SOS AIMD spectrum only exhibits two bands (frequencies shown in the inset of Figure 4B) that have previously been associated with Ag modes of $C_{60}$.[44] Although the isolated gas phase Raman spectrum of $C_{60}$ has not been reported, our computed spectrum is reminiscent of the 784 nm laser driven Raman scattering from $C_{60}$ evaporated on glass.[44] In the prior work, three vibrational resonances at 270, 495, and 1466 $cm^{-1}$ were observed. The absence of the lowest frequency Hg(1) vibration from our SOO-based AIMD

Raman spectrum is likely a result of the relatively short propagation time (4.7 ps total), which prevents the low frequency modes from being adequately sampled. Note that the computed time-averaged mean polarizability of $C_{60}$ is ~ 2093 atomic units or ~310 $Å^3$. This is ~4X larger than the values (~80 $Å^3$) reported for this system using more accurate methods.[45] Nonetheless, the time-varying polarizabilities (albeit about the wrong average value) yield a qualitatively correct spectrum.

**Conclusions**

Altogether, our results suggest that it is possible to gain significant insights into experimental (enhanced) Raman spectra of medium sized as well as relatively large molecular systems using SOO-based AIMD Raman spectral simulations. This is certainly the case for small aromatic thiols which serve as prototypes for enhanced Raman scattering. We also illustrate that the advantages of vibrational spectroscopy (going beyond the static/harmonic/normal mode picture) are preserved using our approach, and that the relative intensities are qualitatively if not quantitatively (at least for the systems considered in this study) correct. As a result of the significantly reduced cost of our approach, we were able to simulate the AIMD-Raman spectra of large systems, including $Ag_{79}$-MBN as well as $C_{60}$.

We conclude by noting that caution should still be exercised in the application of SOO polarizabilities in the AIMD scheme. This is because the approximations invoked in the calculation of SOO polarizabilities may simply prohibit its application, for example, in systems where excited state relaxation is important and governs molecular polarizabilities. Benchmark calculations (akin to the results shown in Figure 1) should always be performed as a starting point. That said, this study paves the way for various applications of AIMD-based Raman spectroscopy to dissect the experimental observables from systems that cannot be tackled using current approaches.

**Figure Captions**

**Figure 1.** Comparison between CPKS and SOO polarizabilities. Solid squares and open circles correspond to the isotropic and anisotropic components of the computed polarizability tensors. The values obtained at the minima of benzene, TP, ATP, NTP, MBN, and DMAB are shown in panel A. Values with the largest deviations from the best linear fit are highlighted, and correspond to the anisotropic components of the polarizabilities of TP and NTP. Time-averaged CPKS and SOO polarizabilities obtained by time averaging the on-the-fly computed values over the entire lengths of the constant energy trajectories for each case are shown in Panel B. Slopes and $R^2$ values from a best linear fit (solid blue lines in A and B) are given in the insets.

**Figure 2.** Comparison between experimental (black lines), static/harmonic Raman spectra evaluated at minimum energy geometries (red lines, sticks individually broadened with Lorentzian functions with 5 cm$^{-1}$ widths), and AIMD Raman spectra obtained through Fourier transforms of CPKS (blue lines) and SOO (green lines) polarizability autocorrelation functions. The results for TP, NTP, and MBN are shown in panels A, B, and C, respectively. Throughout this figure, dashed lines and rectangles are used to highlight relative intensities and spectra regions that are discussed in the main text.

**Figure 3.** Direct comparison between experimental (black lines) and AIMD Raman spectra obtained through Fourier transforms of SOO polarizability autocorrelation functions (green lines). The results for ATP and NTP are shown in panels A and B, whereas the results for DMAB, formed at a plasmonic tip-sample nanojunction are shown in panel (C), and through ESD of ATP in panel (D).

**Figure 4.** Raman spectra of an $Ag_{79}$-MBN complex (A) and $C_{60}$ (B) from AIMD, using the SOO approximation. The experimental (black line) and AIMD Raman (green line) spectra shown in panel A are taken from Figure 2C, for reference, and compared to computed Raman spectrum of the metal-molecule complex (red trace). The insets of the two panels show an overlay of structures (early/intermediate/late time steps in blue/white/red) that together comprise the trajectories, which are 3.4 ps and 4.7 ps long for the $Ag_{79}$-MBN complex (A) and $C_{60}$ (B), respectively.

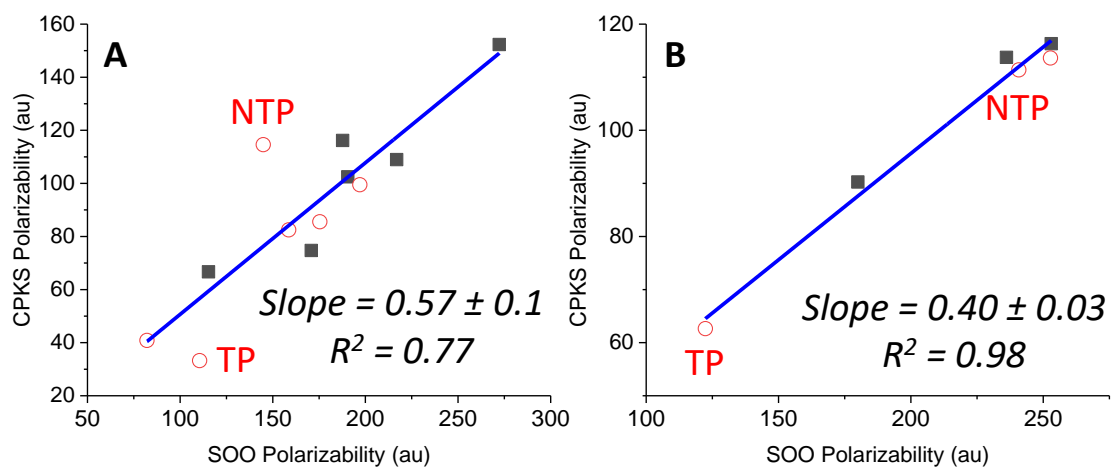

**Figure 1**

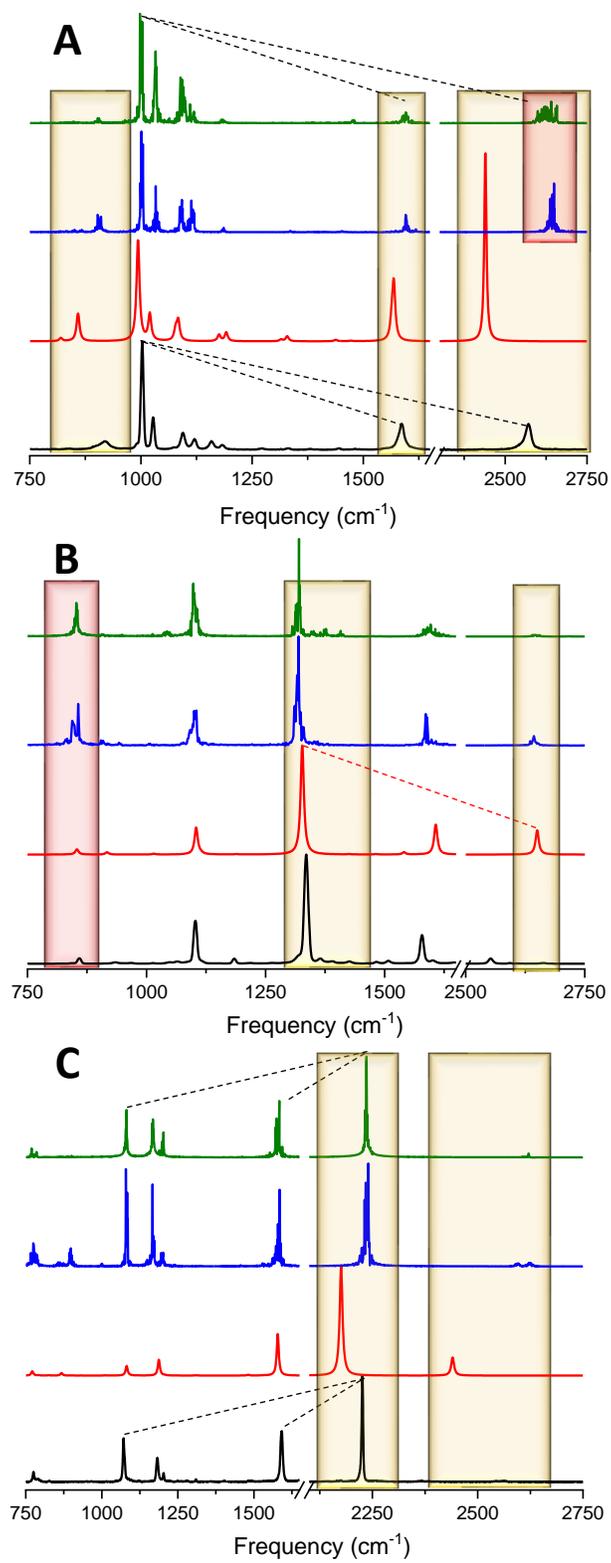

**Figure 2**

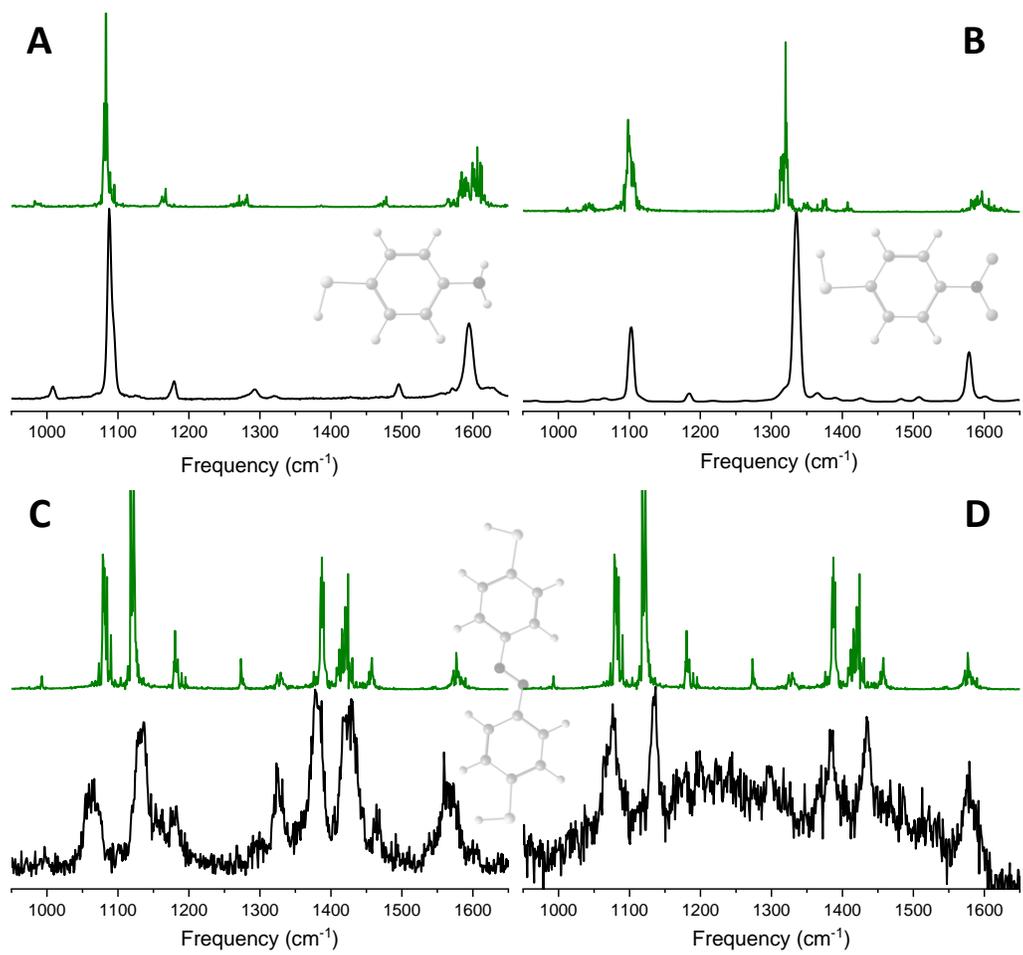

**Figure 3**

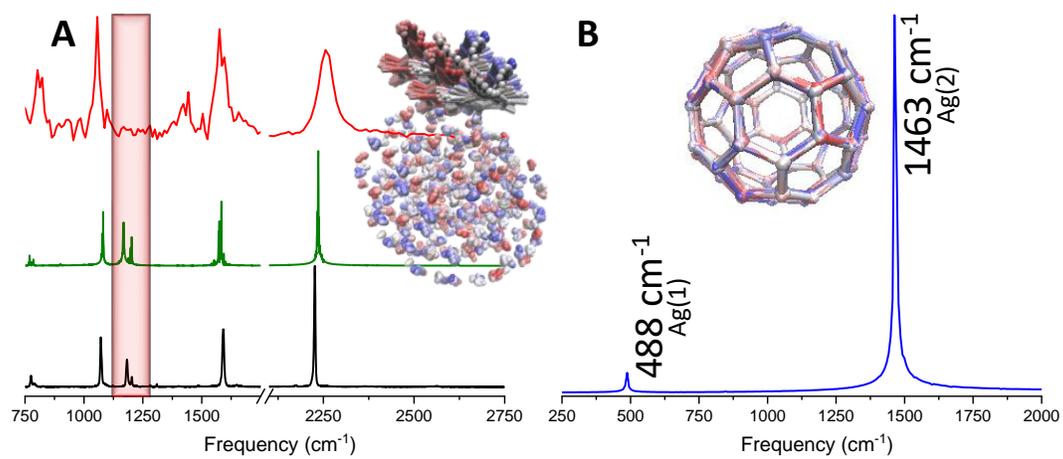

**Figure 4**

**AUTHOR INFORMATION**

**Corresponding Author**

*patrick.elkhoury@pnnl.gov

The authors declare no competing financial interest.

**ACKNOWLEDGMENTS**

AB is supported by the Department of Energy's (DOE) Office of Biological and Environmental Research Bioimaging Technology project #69212. GEJ, EB, NG, and PZE acknowledge support from the US DOE, Office of Science, Office of Basic Energy Sciences, Division of Chemical Sciences, Geosciences & Biosciences. This research benefited from resources provided by PNNL Institutional Computing. This work was performed at EMSL, a DOE Office of Science User Facility sponsored by the Office of Biological and Environmental Research and located at the Pacific Northwest National Laboratory. PNNL is operated by Battelle Memorial Institute for the United States Department of Energy under DOE contract number DE-AC05-76RL1830.

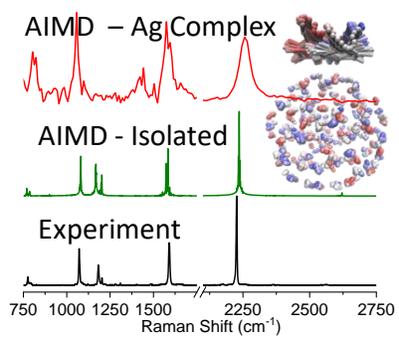

**TOC Graphic**